\documentclass[twocolumn,superscriptaddress,amssymb,amsmath,aps,prl]{revtex4-1}
\usepackage{color}
\usepackage{graphicx}
\usepackage{esint}
\usepackage{mathptmx}
\usepackage{braket}

\makeatletter

\@ifundefined{textcolor}{}
{%
	\definecolor{BLACK}{gray}{0}
	\definecolor{WHITE}{gray}{1}
	\definecolor{RED}{rgb}{1,0,0}
	\definecolor{GREEN}{rgb}{0,1,0}
	\definecolor{BLUE}{rgb}{0,0,1}
	\definecolor{CYAN}{cmyk}{1,0,0,0}
	\definecolor{MAGENTA}{cmyk}{0,1,0,0}
	\definecolor{YELLOW}{cmyk}{0,0,1,0}
}

\@ifundefined{textcolor}{}{%
	\definecolor{BLACK}{gray}{0}
	\definecolor{WHITE}{gray}{1}
	\definecolor{RED}{rgb}{1,0,0}
	\definecolor{GREEN}{rgb}{0,1,0}
	\definecolor{BLUE}{rgb}{0,0,1}
	\definecolor{CYAN}{cmyk}{1,0,0,0}
	\definecolor{MAGENTA}{cmyk}{0,1,0,0}
	\definecolor{YELLOW}{cmyk}{0,0,1,0}
}

\makeatother

\begin{document}
	
	
	\title{Floquet Cavity Electromagnonics}
	
	
	\author{Jing Xu}
	\affiliation{ 
		Center for Nanoscale Materials, Argonne National Laboratory, Lemont, IL 60439, USA
	}
	
	\author{Changchun Zhong}
	\affiliation{ 
		Pritzker School of Molecular Engineering, University of Chicago, Chicago, IL 60637, USA
	}
	
	\author{Xu Han}
	\affiliation{ 
		Center for Nanoscale Materials, Argonne National Laboratory, Lemont, IL 60439, USA
	}
	
	\author{Dafei Jin}
	\affiliation{ 
		Center for Nanoscale Materials, Argonne National Laboratory, Lemont, IL 60439, USA
	}

	\author{Liang Jiang}
	\affiliation{ 
		Pritzker School of Molecular Engineering, University of Chicago, Chicago, IL 60637, USA
	}
	
	\author{Xufeng Zhang}
	\email{xufeng.zhang@anl.gov}
	\affiliation{ 
		Center for Nanoscale Materials, Argonne National Laboratory, Lemont, IL 60439, USA
	}
	
	\date{\today}

	\begin{abstract}
		Hybrid magnonics has recently attracted intensive attentions as a promising platform for coherent information processing. In spite of its rapid development, on-demand control over the interaction of magnons with other information carriers, in particular microwave photons in electromagnonic systems, has been long missing, significantly limiting the broad applications of hybrid magnonics. Here, we show that by introducing Floquet engineering into cavity electromagnonics, coherent control on the magnon-microwave photon coupling can be realized. Leveraging the periodic temporal modulation from a Floquet drive, our first-of-its-kind Floquet cavity electromagnonic system can manipulate the interaction between hybridized cavity electromagnonic modes on demand. Moreover, we demonstrate a new coupling regime in such systems: the Floquet ultrastrong coupling, where the Floquet splitting is comparable with or even larger than the level spacing of the two interacting modes, resulting in the breakdown of the rotating wave approximation. Our findings open up new directions for magnon-based coherent signal processing.
	\end{abstract}
	
	\maketitle


	\textit{Introduction.} Floquet engineering, which refers to temporal modulation of system parameters by periodic drives, has been known as an effective approach for controlling the dynamics of a given system. In recent years, it has been implemented in a large variety of  systems ranging from cold atoms \cite{2017_RMP_Eckardt,2015_NPhysc_Eisert,2007_PRL_Lignier,2011_PRL_Jiang,2017_PRL_Potirniche,2019_PRA_Li} and quantum dots \cite{2016_PRX_Stehlik,2018_PRL_Koski} to  integrated photonics \cite{2019_NPhoton_Zhang} and Josephson junction devices, \cite{2019_PRA_Sameti,2020_NJP_Wang} enabling a diverse variety of novel functionalities. Aside from practical applications, Floquet-driven systems have also significantly advanced fundamental research, leading to the experimental observation of novel non-equilibrium phenomena such as discrete time-crystalline phases \cite{2017_Nature_Zhang,2017_Nature_Choi,2018_PRL_Gong} or Floquet spin-glass phases. \cite{2019_PRL_Raposo} 
	
	Among all coherent information systems, magnonic systems have been emerging as a highly promising platform because of their unique properties. In magnonic systems, magnons---quasiparticles of spin waves---are used as the  information carrier. Their coherent interactions with a broad variety of other systems have been demonstrated recently \cite{2014_PRL_Zhang,2014_PRL_Tabuchi,2014_PRApplied_Goryachev,2015_PRL_Bai,2016_SciAdv_Zhang,2016_PRL_Zhang_Optomagnonics,2016_PRL_Osada,2016_PRL_Haigh,2017_PRB_Sharma,2018_PRB_Graf, 2019_PRL_Hou,2019_PRL_Li}. For instance, magnons, with their frequencies in the gigahertz range, naturally interact with microwave photons through magnetic dipole-dipole interactions. Most importantly, the coupling strength is significantly enhanced by the large spin density in the magnon medium and can reach the strong coupling regime. As of today, such hybrid cavity electromagnonic systems have been experimentally demonstrated in both classical \cite{2014_PRL_Zhang,2014_PRL_Tabuchi,2014_PRApplied_Goryachev,2015_PRL_Bai,2019_PRL_Hou,2019_PRL_Li} and quantum \cite{2015_Science_Tabuchi,2017_SciAdv_Lachance-Quirion,2020_Science_Lachance-Quirion} regimes, becoming the most intensively studied hybrid magnonic systems. With unique magnon properties such as large tunability and time reversal symmetry breaking, novel functionalities can be achieved in these systems \cite{2015_NComm_Zhang,2020_PRAppl_Zhang,2019_PRL_Wang,2019_PRL_Zhang,2020_PRL_Yuan}.
	
	However, unlike in systems such as optomechanics \cite{2008_Nature_Li,2010_Science_Weis,2011_Nature_Safavi-Naeini,2011_Nature_Teufel,2014_RMP_Aspelmeyer} that employ parametric coupling, the direct magnon-photon coupling in a given hybrid electromagnonic system is usually difficult to manipulate. Although magnons naturally possess great tunability, fast or on-demand tuning of the coupling is extremely challenging, making the dynamic control of coherent signals practically impossible. This poses a tremendous obstacle in the broad applications of hybrid electromagnonics. In this work, we show that by introducing Floquet engineering into cavity electromagnonics, in-situ tuning of the magnon-photon interaction is, to the best of our knowledge, achieved for the first time. In our Floquet cavity electromagnonic system, a driving field induces mode splittings that are analogous to the Autler-Townes splitting (ATS) in atomic physics, where the coupling strength between two energy levels is determined by the strength of the Floquet driving field. The system response is studied in both the frequency and temporal domains. More interestingly, our system supports a new coupling regime -- Floquet ultrastrong coupling (FUSC), which has not been observed previously in existing Floquet systems. In this regime, the mode splitting exceeds the energy level spacing of the two interacting modes, distinguishing it from the conventional ultrastrong coupling where the coupling strength is a significant fraction of the bare frequencies of the uncoupled systems. We further show that in this regime, the rotating wave approximation no longer holds and the counter rotating terms in the Hamiltonian start to exhibit non-negligible effects on the system response.	All these findings point to a new direction for advancing magnon-based coherent information processing.

	\begin{figure}[tb]
		\centering
		\includegraphics[width=0.9\linewidth]{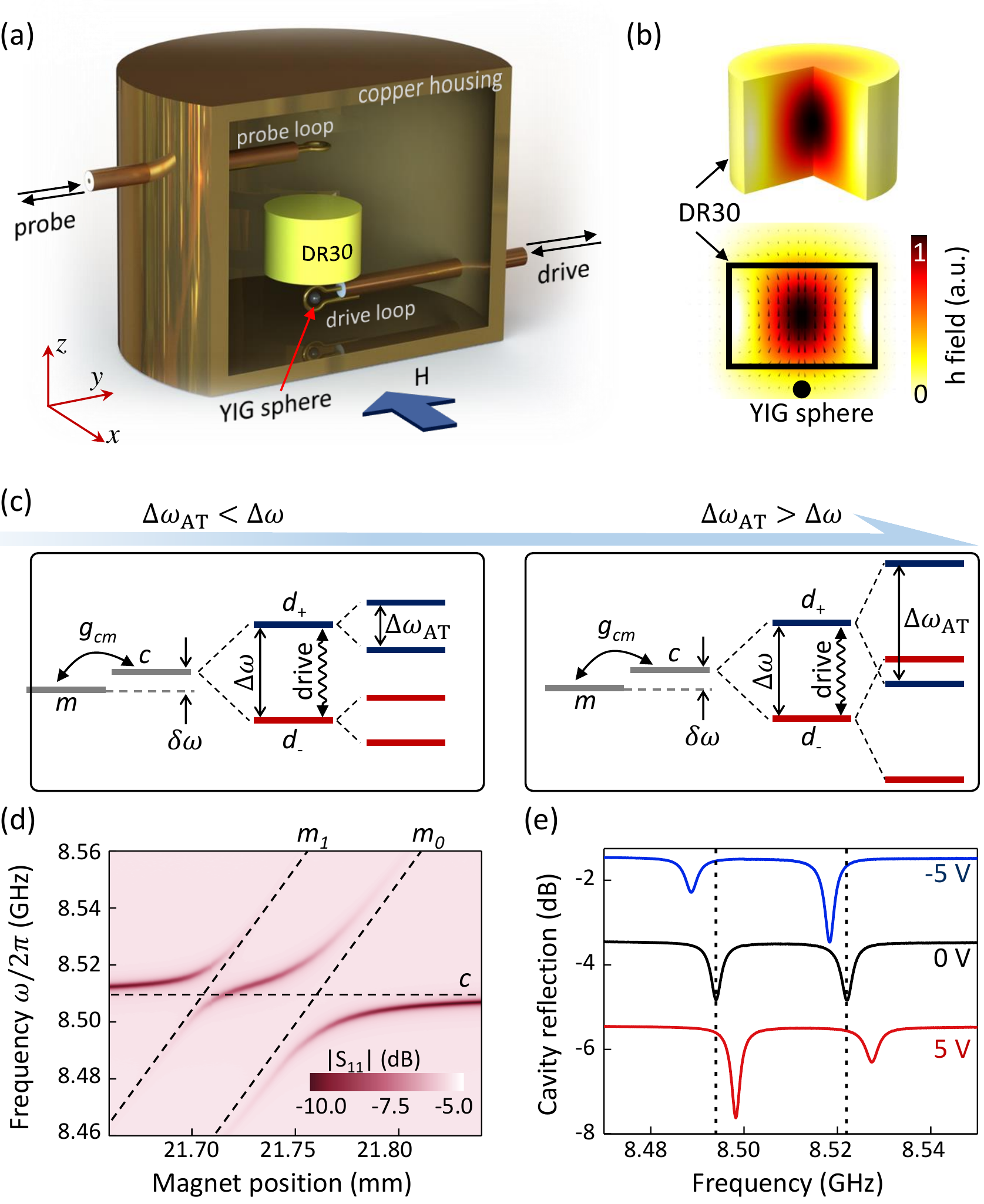}
		\caption{(a) Device schematics. A spherical YIG magnonic resonator is placed underneath a cylindrical dielectric resonator DR30 inside a copper housing. An external magnetic field $H$ is applied along $x$ direction. The coaxial probe loop is for microwave excitation and readout; the drive loop wrapping around the YIG sphere along $x$ provides the Floquet driving field. (b) Simulated magnetic components of the cavity field for the TE$_{01\delta}$ mode of the dielectric resonator. Color: field intensity; arrows: field direction. (c) Energy level diagram. $m$: magnon mode; $c$: microwave photon mode; $\delta\omega$: magnon detuning; $g_{cm}$: magnon-photon coupling strength; $d_\pm$: upper (lower) hybrid mode; $\Delta\omega$: energy separation between two hybrid modes; $\Delta\omega_\mathrm{AT}$: Autler-Townes splitting (ATS). A driving field is applied to enable the transition between hybrid modes $d_-$ and $d_+$. Under strong drives the system enters the Floquet ultrastrong coupling regime where $\Delta\omega_\mathrm{AT}>\Delta\omega$. (d), Measured cavity reflection spectra at various magnetic fields (controlled by the magnet position $x$). Two avoided crossings are visible, corresponding to the strong coupling between the cavity photon mode ($c$) and two magnon modes ($m_0$ and $m_1$). (e), Cavity reflection spectra measured at $x=21.75$ mm with different DC bias conditions applied to the drive loop.}
		\label{fig1}
	\end{figure}

	
	\textit{System Description}. Our system consists of a high-quality dielectric resonator and a highly polished yttrium iron garnet (YIG) sphere [Fig.\,\ref{fig1}(a)]. The cylinder-shaped dielectric resonator (DR30) supports a TE$_{01\delta}$ mode [Fig.\,\ref{fig1}(b)] at 8.5 GHz, with its microwave magnetic fields along the axial ($z$) direction. The cavity mode volume is greatly reduced because of the high dielectric permittivity ($\varepsilon=30$), which enhances its coupling with the magnon mode. The cylinder is hosted inside a copper housing to eliminate radiation losses. Together with the low dielectric loss, this leads to quality factors as high as 10,000 for the cavity resonances at room temperature. A loop antenna is placed above the dielectric resonator with its loop along $z$ direction to probe the cavity photon mode. 
	
	The YIG sphere is placed underneath the dielectric resonator and close to its end surface. With an external bias magnetic field that saturates all the spins, it supports a uniform magnon mode (the Kittel mode) whose frequency is determined by the field strength. When the bias field is not parallel with the microwave magnetic fields of the photon mode, there exists non-zero magnon-photon interaction. In order to maximize such interaction, the bias field is applied perpendicular to $z$ (along $x$ direction). In our experiments, the bias field strength is controlled by the magnet position along $x$ direction.

	The system is described by the Hamiltonian:
	\begin{equation}\label{eq_full}
	\hat{H}=\hbar\omega_c\hat{c}^\dagger\hat{c}+\hbar\omega_m\hat{m}^\dagger\hat{m}+\hbar g_{cm}(\hat{c}^\dagger\hat{m}+\hat{c}\hat{m}^\dagger)+\hat{H}_\mathrm{F},
	\end{equation}
	\noindent where $\hbar$ is the reduced Planck's constant, $\hat{c}^\dagger$ and $\hat{c}$  ($\hat{m}^\dagger$ and $\hat{m}$) are the creation and annihilation operators for the cavity photon (magnon) mode, $g_{cm}$ is the beam-splitter type coupling strength, $\omega_c$ ($\omega_m$) is the resonance frequency of the cavity photon (magnon) mode, and $\hat{H}_\mathrm{F}$ is the Floquet driving term. The magnon frequency is controlled by an external magnetic field: $\omega_m=\gamma H$, where $\gamma=28$ GHz/T is the gyromagnetic ratio, and $H$ is the bias magnetic field. Under strong coupling condition $2g_{cm}>\kappa_c,\kappa_m$ where $\kappa_c$ and $\kappa_m$ represent the dissipation rate of the cavity photon and magnon mode, respectively, the two modes hybridize with each other, allowing coherent information conversion among the magnonic and electromagnetic degrees of freedom. In our experiment, strong coupling is confirmed by the avoided crossing features in the measured cavity reflection spectra when the magnon frequency is swept by varying the magnet position [Fig.\,\ref{fig1}(d)].

	The Floquet driving is realized in our system through frequency modulation of the magnon mode. A small coil, which has previously been used only for GHz readout or controls \cite{2019_NJP_Boventer}, is looped tightly around the YIG sphere as a means to modulate the bias magnetic field.  The loop is aligned along the bias field ($x$) direction, and has only 3 turns in order to reduce the inductance to allow fast modulation. The effect of the driving coil is confirmed by the shifted magnon resonances when different DC drives are applied [Fig.\,\ref{fig1}(e)]. With a sinusoidal drive, the Floquet term of the Hamiltonian reads:	\begin{equation}\label{eq_Floquet}
	\hat{H}_\mathrm{F}=\hbar\Omega\hat{m}^\dagger\hat{m}\cos(\omega_\mathrm{D} t),
	\end{equation}
	\noindent where $\Omega$ and $\omega_\mathrm{D}$ are the strength and frequency of the driving field, respectively.

	\begin{figure}[t]
		\centering
		\includegraphics[width=0.95\linewidth]{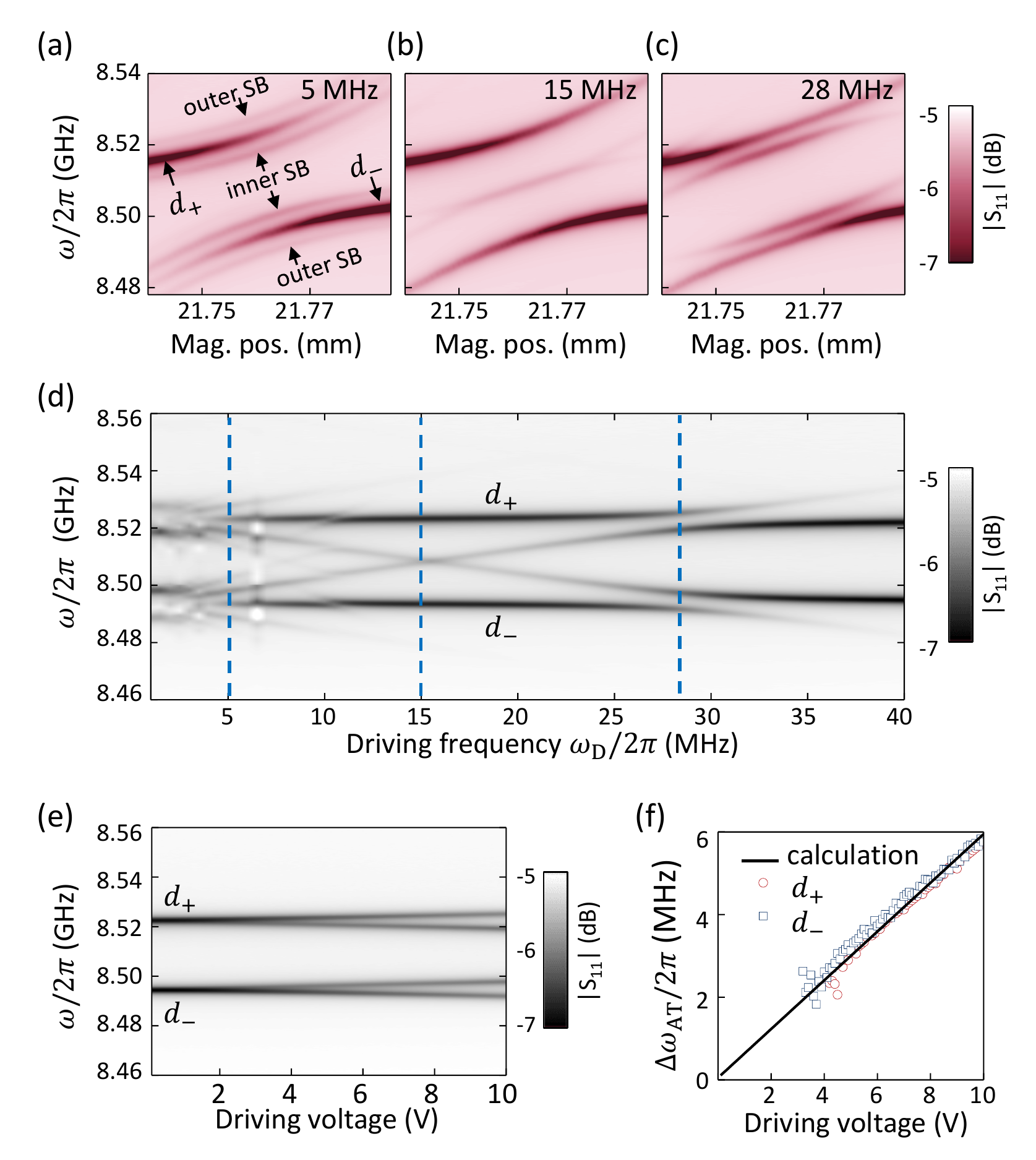}
		\caption{(a)--(c), Measured cavity reflection spectra versus the bias magnetic field at a 10-V peak-to-peak driving amplitude for three frequencies: $\omega_\mathrm{D}=5,15,28$ MHz, respectively. In addition to the avoided crossing caused by the strong coupling between the cavity photon mode ($c$) and magnon mode ($m_0$), sidebands are created by the AC driving fields. (d), Measured cavity reflection spectra as a function driving frequency at $x=21.762$ mm when $\delta\omega=0$. The dashed lines correspond to conditions for (a)--(c). (e), Measured cavity reflection spectra as a function of driving amplitude, with $\delta\omega=0$ and $\omega_\mathrm{D}=28$ MHz. (f), Extracted Autler-Townes splitting for both $d_-$ (square) $d_+$ (circle) modes in (e) as a function of the driving voltage. Solid line is from numerical calculation. SB: sideband.}
		\label{fig2}
	\end{figure}

	\textit{Magnonic Autler-Townes Effect}. When the magnon and photon modes are on resonance ($\delta\omega=\omega_m-\omega_c=0$), two hybrid modes $d_\pm=(\hat{c}\pm\hat{m})/\sqrt{2}$ form at frequencies $\omega_\pm=\omega_c\pm g_{cm}$. Using $d_\pm$ as the new basis and under rotating wave approximation (RWA), the Hamiltonian can be rewritten as
	\begin{equation}\label{eq_RWA}
	\hat{H}_\mathrm{RWA}=\frac{\delta\omega_\mathrm{D}}{2}\hat{d}^\dagger_{-}\hat{d}_{-}-\frac{\delta\omega_\mathrm{D}}{2}\hat{d}_{+}^\dagger\hat{d}_{+}+\frac{\Omega}{4}(\hat{d}_{-}^\dagger\hat{d}_{+}+\hat{d}_{-}\hat{d}_{+}^\dagger),
	\end{equation}
	\noindent where $\delta\omega_\mathrm{D}=\Delta\omega-\omega_\mathrm{D}$. Here $\Delta\omega=\omega_+-\omega_-=2g_{cm}$ is the level spacing between the two hybrid modes. This Hamiltonian shows that the external Floquet field drives the transition between the two hybrid modes, which is analogous to a two-level system driven by a laser field. As a result, each hybrid mode splits further into two energy levels with a level spacing $\Delta\omega_\mathrm{AT}$ that is determined by the driving strength, resembling the ATS in a laser-driven two-level system  [Fig.\,\ref{fig1}(c)].

	The cavity responses under a Floquet drive are plotted in Figs.\,\ref{fig2}\,(a)--(c), which are zoomed in to show the avoided-crossing between the fundamental magnon mode ($m_0$) and the cavity mode ($c$). Two hybrid modes ($d_\pm$) resulting from the magnon-photon hybridization are clearly visible.
	The Floquet drive creates two sidebands for each hybrid mode. As the driving frequency $\omega_\mathrm{D}$ increases, the inner sideband of one hybrid mode moves closer to, and eventually meets with, the other hybrid mode (when $\omega_\mathrm{D}=\Delta\omega=2\pi\times28$ MHz), leading to the ATS discussed above [Fig.\,\ref{fig2}(c)].

	\begin{figure}[t]
		\centering
		\includegraphics[width=0.95\linewidth]{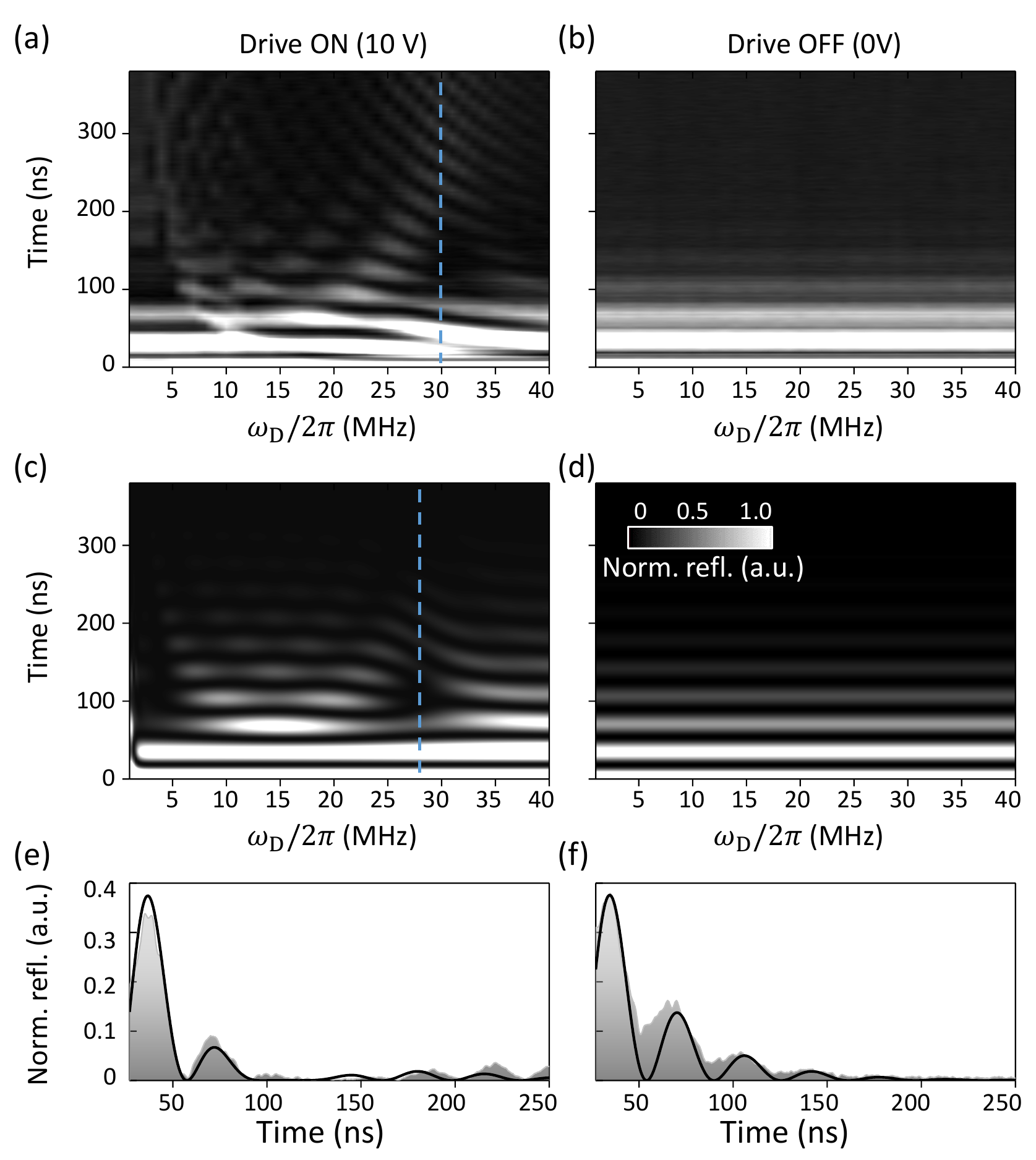}
		\caption{(a),(b) Measured cavity temporal response under different driving frequencies after a pulsed excitation (center frequency: 8.52 GHz, width: 20 ns) at time zero when the Floquet drive is on and off, respectively. 
			(c),(d) Calculated cavity temporal response with and without the Floquet drive, respectively. (e),(f) Line plot of the theoretical (solid line) and experimental (greyed area) temporal responses at Floquet driving frequencies as indicated by the vertical dashed lines in (a)--(d) ($\omega_\mathrm{D}=2\pi\times 28$ MHz for the theoretical calculation, while $\omega_\mathrm{D}=2\pi\times 31$ MHz for experiment results which is slightly higher because of magnon non-linearity effects induced by the strong microwave input) with and without the Floquet drive, respectively. Norm. refl.: normalized reflection.}
		\label{fig3}
	\end{figure}

	The on-resonance ($\delta\omega=0$) response of the system as a function of the driving frequency is shown in Fig.\,\ref{fig2}(d). The most prominent feature is that the upper (lower) sideband of $d_-$ ($d_+$) mode moves closer to the other mode $d_+$ ($d_-$) as the driving frequency increases, and eventually crosses that mode with an avoided-crossing feature at $\omega_\mathrm{D}=\Delta\omega=2\pi\times28$ MHz, which corresponds to the ATS. Note that our modulation approach creates sidebands at both $\omega_\pm\pm\omega_\mathrm{D}$ frequencies, therefore there also exists a lower (upper) sideband for $d_-$ ($d_+$), which rapidly disappears as it moves away from the hybrid mode because of the lacking of coupling with the other mode.
	
	Analogous to the laser-driven two-level systems, the ATS observed in our system is also determined by the strength of the driving field. The driving strength, characterized by $\Omega$, is controlled by the signal voltage sent into the driving coil in our experiments. Larger driving strengths result in wider ATS, as shown in Fig.\,\ref{fig2}(e), where the on-resonance cavity reflection is plotted as a function of the driving voltage at a driving frequency $\omega_\mathrm{D}=2\pi\times 28$ MHz. According to Eq.\,(\ref{eq_RWA}), the ATS is linearly proportional to the driving strength $\Delta\omega_\mathrm{AT}=\Omega/2$. This is confirmed by Fig.\,\ref{fig2}(f), where the extracted and calculated ATS show an excellent agreement with each other.

	In addition to modifying the equilibrium spectra, the Floquet drive also provides a new approach for manipulating the electromagnonic dynamics. This has long been a grand challenge such that all previous demonstrations were based on quasi-static controls \cite{2014_PRL_Zhang, 2017_NComm_You, 2019_PRL_Zhang}. 
	Figures\,\ref{fig3}(a) and (c) plot the measured and calculated cavity reflection signals after a 20-ns-wide microwave pulse centered at 8.52 GHz is sent into the cavity, which show a great agreement. At low driving frequencies where the Floquet drive cannot induce coupling between the two hybrid modes $d_\pm$, periodic ``Rabi-like'' oscillations between the magnon and cavity photon modes are observed with a period $T=2\pi/2g_{cm}=36$ ns. This is more clearly visible when the Floquet drive is completely turned off [Figs.\,\ref{fig3}(b) and (d)], and it is evident that the amplitude of the oscillating signal monotonically decreases as a result of dissipation [Figs.\,\ref{fig3}(f)]. When the Floquet drive is turned on and, in particular, when the driving frequency matches the spacing between the two hybrid modes ($\omega_\mathrm{D}=\omega_+-\omega_-= 2\pi\times 28$ MHz), coherent coupling between the hybrid modes is enabled and consequently the temporal response of the system is substantially modified. As shown in Fig.\,\ref{fig3}(e), the oscillation amplitude first rapidly decays into a minimum at around 120 ns and then increases again at around 200 ns, with the time interval matching the ATS ($2\pi/2\Delta\omega_\mathrm{AT}\approx80$ ns for a 10V Floquet drive). The measurement results agree well with our theoretical model. Because such Floquet drive-induced coherent interaction can be controlled by the amplitude of the drive, it provides opportunities for complex real-time manipulations of the magnon-photon coupling using electrical pulses.

	
	\textit{Floquet Ultrastrong Coupling}. In the above analysis, the Floquet drive is relatively weak, yielding  small ATS $\Delta\omega_\mathrm{AT}<\Delta\omega$. However, in the strong-drive regime where the ATS is comparable with or even larger than the level spacing between $d_\pm$ ($\Delta\omega_\mathrm{AT}\gtrapprox\Delta\omega$), the RWA Hamiltonian in Eq.\,(\ref{eq_RWA}) is no longer sufficient to describe the system. This corresponds to a novel and rarely investigated coupling regime: the FUSC regime. Here, the counter rotating term need to be included and the Hamiltonian becomes:
	\begin{equation}\label{Eq_Bessel}
	\begin{aligned}
	\hat{H}={}& \omega_{-}\hat{d}_{-}^\dagger\hat{d}_{-}+\omega_{+}\hat{d}_{+}^\dagger\hat{d}_{+}+\\
	& g_{cm}\sum_{n=\text{odd}}^{\infty}J_n(\frac{\Omega}{\omega_\mathrm{D}})(\hat{d}_{-}^\dagger\hat{d}_{+}e^{i(n\omega_\mathrm{D} t)}+\hat{d}_{-}\hat{d}_{+}^\dagger e^{-i(n\omega_\mathrm{D} t)}).
	\end{aligned}
	\end{equation}
	\noindent where $n$ is the sideband order, and $J_n(x)$ is the $n$-th Bessel function of the first kind. 
	
	\begin{figure}[tb]
		\centering
		\includegraphics[width=\linewidth]{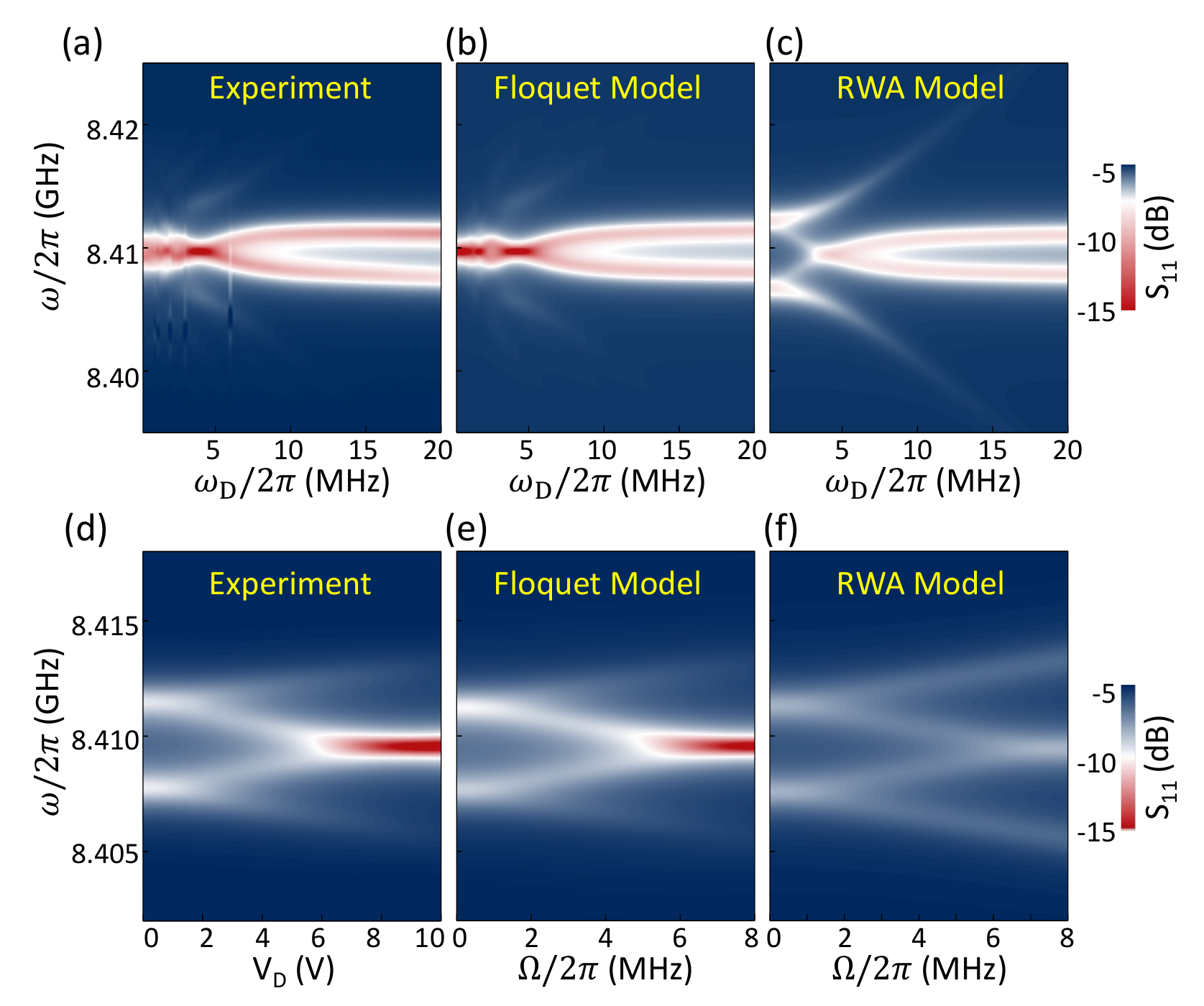}
		\caption{(a)--(c), Cavity reflection spectra when the driving frequency $\omega_\mathrm{D}$ is swept at a fixed driving amplitude (10 V peak-to-peak), obtained from experiment, Floquet model, and rotating wave approximation (RWA) model, respectively. (d)--(f), Cavity reflection spectra when the driving amplitude is swept at a fixed driving frequency ($\omega_\mathrm{D}=2\pi\times 3.85$ MHz), obtained from experiment, Floquet model, and RWA model, respectively.}
		\label{fig4}
	\end{figure}

	The last term in Eq.\,(\ref{Eq_Bessel}) represents the summation of interactions between different sidebands of $d_\pm$ modes. The Floquet drive generates a series of sidebands at frequencies $\omega_\pm\pm n\omega_\mathrm{D}$. Multiple sidebands and their interactions with the other mode are clearly visible in Fig.\,\ref{fig2}(d) when $\omega_\mathrm{D}$ is small. Specifically, the individual coupling strength of the $n$-th sideband of mode $d_\pm$ with mode $d_\mp$ is determined as $g_n=g_{cm}J_n(\frac{\Omega}{\omega_\mathrm{D}})$. Apparently, this coupling strength is determined by the intrinsic magnono-photon coupling strength $g_{cm}$, but can be controlled by the  frequency and strength of the Floquet drive. Interestingly, only sidebands with odd $n$ values have non-zero coupling strengths. This is confirmed by the experimental observation in Fig.\,\ref{fig2}(d), where only the first and third sidebands of $d_\pm$ cross $d_\mp$ with an avoided crossing (at 28 and 11 MHz, respectively), while the second sideband does not. When the Floquet drive is relatively small, the ATS can be determined as $\Delta\omega_\mathrm{AT}=2g_n$. But this does not apply for strong drives, where the effects from multiple sidebands need to be considered.

	In general, experimental investigation of the strong-drive regime is very challenging because the maximally achievable driving strength is limited. Alternatively, it can be achieved by reducing the level spacing $\Delta\omega$ to make it comparable with or smaller than the ATS $\Delta\omega_\mathrm{AT}$. This is realized with a reduced intrinsic magnon-photon coupling strength $g_{cm}$ by increasing the gap between the YIG sphere and the dielectric resonator surface.
	Figure\,\ref{fig4}(a) shows the measured cavity reflection spectra as a function of the driving frequency with a much reduced magnon-photon coupling strength $g_{cm}=2\pi\times1.825$ MHz, where ATS is observed at $\omega_\mathrm{D}=2g_{cm}=2\pi\times3.650$ MHz for both $d_-$ and $d_+$. This agrees well with the theory results calculated using the Floquet scattering matrix derived from the full Hamiltonian Eq.\,(\ref{eq_full}), as shown in Fig.\,\ref{fig4}(b). As a comparison, the calculated cavity reflection spectra based on  the Hamiltonian with RWA in Eq.\,(\ref{eq_RWA}) are plotted in Fig.\,\ref{fig4}(c), which exhibit severe deviation from the experimental results, showing the significant effects of the counter rotating term.

	Similarly, the measured cavity reflection spectra under different driving strengths are plotted in Fig.\,\ref{fig4}(d), which show excellent agreement with theory calculations based on the full Floquet Hamiltonian [Fig.\,\ref{fig4}(e)], as well as large deviation from the calculations using RWA [Fig.\,\ref{fig4}(f)]. Compared with the RWA results, the RWA breaking has two major effects: first, the outer two branches become much weaker than the inner two branches in both $\omega_\mathrm{D}$-dependent [Figs.\,\ref{fig4}\,(a)--(c)] and $\Omega$-dependent [Figs.\,\ref{fig4}\,(d)--(f)] cavity reflections; second, the inner two branches merge and form a much deeper dip.

	\textit{Conclusion}. To conclude, this work demonstrates a Floquet cavity electromagnonic system that provides controllable hybridization between magnons and microwave photons. The Floquet engineering technique provides a versatile approach for manipulating hybrid magnon states and enables the observation of on-demand ATS. The measurement results in both the frequency and temporal domains show excellent agreements with our Floquet driving model and the corresponding scattering matrix. More importantly, this approach leads to a new coupling regime---FUSC---for hybrid magnonics, where RWA breaks down and new phenomena are observed. This technique can be readily applied as a general approach to a broad range of magnonic systems that have been lacking the essential controllability, producing not only new functionalities but also novel non-equilibrium magnon dynamics. For instance, more complicated spectral control can be obtained if the driving strength can be further enhanced. Besides, time domain operations such as Rabi or Ramsey pulse sequences will allow on-demand mode swapping or storage. Although our experiments are carried out at room temperatures in the classical regime, the principles demonstrated here also directly apply to quantum operations, where hybrid magnonics has shown great potentials for applications such as quantum transduction.

	\begin{acknowledgments}
		This work was performed, in part, at the Center for Nanoscale Materials, a U.S. Department of Energy Office of Science User Facility, and supported by the U.S. Department of Energy, Office of Science, under Contract No. DE-AC02-06CH11357. L.J. acknowledges support from the ARL-CDQI (W911NF-15-2-0067), ARO (W911NF-18-1-0020, W911NF-18-1-0212), ARO MURI (W911NF-16-1-0349), AFOSR MURI (FA9550-15-1-0015, FA9550-19-1-0399), NSF (EFMA-1640959, OMA-1936118), and the Packard Foundation (2013-39273).
	\end{acknowledgments}

\section*{Appendix I: The Floquet scattering matrix}
The system is described by the following periodic Hamiltonian ($
\hbar=1$)
\begin{equation}\label{eq1}
\begin{split}
    \hat{H}&=\omega_c\hat{c}^\dagger\hat{c}+\omega_m\hat{m}^\dagger\hat{m}+g_{cm}\hat{c}^\dagger\hat{m}+g_{cm}^\ast\hat{c}\hat{m}^\dagger \\ &+\Omega\hat{m}^\dagger\hat{m}\cos(\omega_\mathrm{D} t+\phi),
\end{split}
\end{equation}
where $\hat{c},\hat{m}$ are the microwave and magnon mode operators, with corresponding mode frequencies $\omega_c$ and $\omega_m$. $g_{cm}$ is the Beam-splitter type coupling strength. $\Omega$ and $\omega_\mathrm{D}$ are the drive strength and drive frequency, respectively. In the Heisenberg picture, the mode operators follow the dynamics $\dot{\textbf{A}}(t)=-i\textbf{H}(t)\textbf{A}(t)$, which is short for
\begin{equation}
    \frac{d}{dt}\begin{pmatrix}\hat{c}(t)\\ \hat{m}(t)\end{pmatrix}
    =-i\begin{pmatrix}
    \omega_c&g_{cm}\\g_{cm}^\ast&\omega_m+\Omega\cos(\omega_\mathrm{D}t+\phi)
    \end{pmatrix}\begin{pmatrix}\hat{c}(t)\\ \hat{m}(t)\end{pmatrix}.
\end{equation}
It is a periodic time dependent differential equation. According to Floquet theorem, the eigen-mode can be written in the form $\textbf{A}_m(t)=\textbf{a}_m(t)e^{-i{\epsilon_m} t}$, where $\epsilon_m$ is the called the Floquet quasi-energy. $\textbf{a}_m(t)$ is periodic in time with the same period as $\textbf{H}(t)$ and obviously satisfies the eigen equation
\begin{equation}\label{eq3}
    \left[\textbf{H}(t)-i\frac{d}{dt}\right]\textbf{a}_m(t)=\epsilon_m\textbf{a}_m(t).
\end{equation}
In order to find the Floquet mode, we write the time-dependent mode in its Fourier component $\textbf{a}_m(t)=\sum_{N}\textbf{a}_m^Ne^{-iN\omega_\mathrm{D}t}$. If we further take the Fourier index as a new mode index, the above time-dependent differential equation can be mapped to a time-independent tight-binding Hamiltonian $\textbf{H}_F$ \cite{Shirley1965}, with matrix elements given by
\begin{equation}
    \braket{\alpha n|H_F|\gamma k}=H_{\alpha\gamma}^{n-k}+n\omega_\mathrm{D}\delta_{{\omega_\mathrm{D}}\alpha\gamma}\delta_{{\omega_\mathrm{D}}nk},
\end{equation}
where $H_{\alpha\gamma}$ is the matrix element of $\textbf{H}(t)$ and it has the Fouirer expansion $H_{\alpha\gamma}=\sum_nH^n_{\alpha\gamma}e^{-in\omega_\mathrm{D}t}$. The matrix $\textbf{H}_F$ is usually called the Floquet Hamiltonian. The problem of solving the time-dependent system is then reduced to find the eigenvalue of the equation
\begin{equation}
    \textbf{H}_F\textbf{V}=\epsilon_m \textbf{V},
\end{equation}
where the vector $\textbf{V}=\{...,\hat{c}_{\omega_c-2\omega_\mathrm{D}},\hat{c}_{\omega_c-\omega_\mathrm{D}},\hat{c}_{\omega_c},\hat{c}_{\omega_c+\omega_\mathrm{D}},\\ \hat{c}_{\omega_c+2\omega_\mathrm{D}},...,\hat{m}_{\omega_m-2\omega_\mathrm{D}},\hat{m}_{\omega_m-\omega_\mathrm{D}},\hat{m}_{\omega_m},\hat{m}_{\omega_m+\omega_\mathrm{D}},\hat{m}_{\omega_m+2\omega_\mathrm{D}},...\}^T$. The equation is infinitely dimensional and is generally numerically solved by truncating the matrix.

The above discussion ignores the system losses and the readout. To get the scattering matrix, we need to take them into account. Suppose the magnon mode suffers from loss with rate $\kappa_m$, the cavity has readout rate $\kappa_{c,ex}$, intrinsic loss $\kappa_{c,i}$ and $\kappa_c=\kappa_{c,ex}+\kappa_{c,i}$. Thus each mode follows the dynamical equation
\begin{equation}\label{deq}
    \frac{d}{dt}\textbf{V}=-i\textbf{H}_F\textbf{V}+\textbf{D}\textbf{V}+\textbf{M}\textbf{V}_{in}
\end{equation}
where $\textbf{D}=\text{Diag}[\kappa_c/2,\kappa_m/2]\otimes \textbf{I}_d$ and $\textbf{I}_d$ is an identity matrix with dimension $d$ determined by the truncation. The matrix $\textbf{M}=\text{Diag}[\textbf{I}_d\otimes\begin{pmatrix}\sqrt{\kappa_{c,ex}}&\sqrt{\kappa_{c,i}}\end{pmatrix},\textbf{I}_d*\sqrt{\kappa_m}]$. $\textbf{V}_{in}$ collects all the input operators. The Floquet scattering matrix can be obtained by combining the input-output theory
\begin{equation}
    \textbf{V}_{out}=\textbf{M}^T\textbf{V}-\textbf{V}_{in}.
\end{equation}
The Floquet scattering matrix is given
\begin{equation}\label{eq8}
    \textbf{S}_F[\omega]=\textbf{M}^T(-i\omega \textbf{I}_{2d}+i\textbf{H}_F-\textbf{D})^{-1}\textbf{M}-\textbf{I}_{3d}.
\end{equation}
In the numerical calculation, the element of $d_{th}$ row and $d_{th}$ column gives the reflection from the cavity, which we denoted as $S_{11}$. In general, the analytical expression of $S_{11}$ is very complicated resulting from the matrix inverse in Eq. \ref{eq8}. To get a simplified expression, before calculating the matrix inverse in Eq. \ref{eq8}, we split it into diagonal and anti-diagonal part, then expand it by adoping the Woodbury identity \cite{footnote1}. By collecting the first several terms, we arrive at     
\begin{equation}
\begin{split}
    &S_{11}=-1+\frac{\kappa_{c,ex}}{\kappa_c/2-i(\omega-\omega_c)}\times\\ 
    &\left(1-
    \frac{g^2_{cm}}{\kappa_c/2-i(\omega-\omega_c)}\sum_{n=-d}^d\frac{J^2_n(\frac{\Omega}{\omega_\mathrm{D}})}{\kappa_m/2-i(\omega+n\omega_\mathrm{D})+i\omega_m}+... \right),
    \end{split}
\end{equation}
where $J_n(\Omega/\omega_\mathrm{D})$ is the $n_{th}$ Bessel function of the first kind. {We see there is no phase $\phi$ dependence in the reflection spectrum and we will set them to be zero in later discussions.}

With the dynamical equation Eq.(\ref{deq}), one can further evaluate the response signal in the time domain. In the numerical calculation, we suppose the cavity mode is populated by the input pulse with $\Bar{n}$ photons, which will be flopping back and forth among different modes. Thus, we have 
\begin{equation}
    \frac{d}{dt}\textbf{V}=(-i\textbf{H}_F+\textbf{D})\textbf{V},
\end{equation}
with an initial condition $\textbf{V}_{t=0}=\{0,0,...\sqrt{\Bar{n}}_{\omega_c},...,0,0\}^T$. This first order differential can be easily solved by diagonalizing the coefficient matrix. The output signal in the time domain is giving by $c_{\omega_c}^\text{out}(t)=\sqrt{\kappa_{c,ex}}c_{\omega_c}(t)$.

\section*{Appendix II: The modified coupling strength}

In the Floquet picture, we know different side-bands are going to show up due to the periodic drive. The driving field will also modify the coupling strength, which can be seen by analyzing the Hamiltonian
\begin{equation}
    \hat{H}=\omega_c\hat{c}^\dagger\hat{c}+\omega_m\hat{m}^\dagger\hat{m}+g_{cm}(\hat{c}^\dagger\hat{m}+\hat{c}\hat{m}^\dagger)
    +\Omega\hat{m}^\dagger\hat{m}\cos(\omega_\mathrm{D} t).
\end{equation}
Define a dynamical phase
\begin{equation}
\begin{split}
    \varepsilon(t)=\frac{\Omega}{\omega_\mathrm{D}}\left[\sin{(\omega_\mathrm{D} t)}\right],
\end{split}
\end{equation}
then express the Hamiltonian in the rotating frame $U=\exp[i\varepsilon(t)\hat{m}^\dagger\hat{m}]$,
\begin{equation}\label{eq11}
    \hat{H}=\omega_c\hat{c}^\dagger\hat{c}+\omega_m\hat{m}^\dagger\hat{m}+g_{cm}(\hat{c}^\dagger\hat{m}e^{-i\varepsilon(t)}+\hat{c}\hat{m}^\dagger e^{i\varepsilon(t)}).
\end{equation}
The phase factor $\varphi(t)=e^{i\varepsilon(t)}$ can be expressed in the following series
\begin{equation}
    \varphi(t)=\sum_{n=-\infty}^\infty J_n(\frac{\Omega}{\omega_\mathrm{D}}) e^{-i\cdot n(\omega_\mathrm{D} t)},
\end{equation}
where $J_n(\Omega/\omega_\mathrm{D})$ is the $n_{th}$ Bessel function of the first kind with argument $\Omega/\omega_\mathrm{D}$. Thus we have 
\begin{equation}
\begin{split}
    \hat{H}&=\omega_c\hat{c}^\dagger\hat{c}+\omega_m\hat{m}^\dagger\hat{m}\\
    &+g_{cm}\sum_{n=-\infty}^\infty J_n(\frac{\Omega}{\omega_\mathrm{D}})(\hat{c}^\dagger\hat{m}e^{i\cdot n(\omega_\mathrm{D} t)}+\hat{c}\hat{m}^\dagger e^{-i\cdot n(\omega_\mathrm{D} t)} ).
\end{split}
\end{equation}
From the above Hamiltonian, we see that the coupling strength is modified by the Bessel function. Specifically, if we look at the term with $n=0$, we have
\begin{equation}
\begin{split}
    \hat{H}&=\omega_c\hat{c}^\dagger\hat{c}+\omega_m\hat{m}^\dagger\hat{m}+g_{cm}J_0(\frac{\Omega}{\omega_\mathrm{D}})(\hat{c}^\dagger\hat{m}+\hat{c}\hat{m}^\dagger)
    \\
    &+g_{cm}\sum_{n\neq 0} J_n(\frac{\Omega}{\omega_\mathrm{D}})(\hat{c}^\dagger\hat{m}e^{-i\cdot n(\omega_\mathrm{D} t)}+\hat{c}\hat{m}^\dagger e^{i\cdot n(\omega_\mathrm{D} t)} ),
\end{split}
\end{equation}
where a modified coupling strength $g_{cm}J_0(\Omega/\omega_\mathrm{D})$ between the magnon and the cavity mode is obtained.

\section*{Appendix III: Crossing and anti-crossing of reflection spectrum}

In the main text, we see in the reflection spectrum that some side-bands are coupled and others are not when we tune the drive frequency.
Since the magnon and cavity mode are strongly coupled,approximately two hybridized modes can be defined (this is true in the strong coupling limit $g_{cm}\gg\kappa_c,\kappa_m$ and take $\omega_c=\omega_m=\omega_0$)
\begin{equation}
\begin{split}
    \hat{d}_+&=\frac{\sqrt{2}}{2}(\hat{c}+\hat{m}),\\
    \hat{d}_-&=\frac{\sqrt{2}}{2}(\hat{c}-\hat{m}),
\end{split}
\end{equation}
with which we can rewrite the Hamiltonian Eq. \ref{eq11},
\begin{equation}
    \hat{H}=\omega_+\hat{d}_+^\dagger\hat{d}_++\omega_-\hat{d}_-^\dagger\hat{d}_-+g_{cm}\sin\varepsilon(t)(i\hat{d}_+^\dagger\hat{d}_--i\hat{d}_+\hat{d}_-^\dagger),
\end{equation}
where $\omega_+=\omega_0+g_{cm}\cos\varepsilon(t)$, $\omega_-=\omega_0-g_{cm}\cos\varepsilon(t)$. We see the hybridized mode separation is given by $2g_{cm}\cos\varepsilon(t)$. A time integral of the separation will give $2g_{cm}J_0(\Omega/\omega_\mathrm{D})$, where we see the separation is approximated by $2g_{cm}$ when either $\Omega/\omega_D$ approaches zero or infinity. Making use of the identity
\begin{equation}
    \sin(z\sin \theta)=-i\sum_{n=\text{odd}}^{\infty}J_n(z)e^{in\theta},
\end{equation}
we have
\begin{equation}
\begin{split}
    \hat{H}&=\omega_+\hat{d}_+^\dagger\hat{d}_++\omega_-\hat{d}_-^\dagger\hat{d}_-\\
    &+g_{cm}\sum_{n=\text{odd}}^{\infty}J_n(\frac{\Omega}{\omega_\mathrm{D}})(\hat{d}_+^\dagger\hat{d}_-e^{in(\omega_\mathrm{D} t)}+\hat{d}_+\hat{d}_-^\dagger e^{-in(\omega_\mathrm{D} t)}).
    \end{split}
\end{equation}
It is important to notice that in the above expression, we only have odd terms in the summation. Based on this Hamiltonian, we can look at the coupled side-bands by taking the corresponding rotating frame transformation. For example,
$U=e^{i\omega_\mathrm{D}\hat{d}_+^\dagger\hat{d}_+t}$, we get 
\begin{equation}
\begin{split}
    \hat{H}&=(\omega_+-\omega_\mathrm{D})\hat{d}_+^\dagger\hat{d}_++\omega_-\hat{d}_-^\dagger\hat{d}_-\\
    &+g_{cm}J_1(\frac{\Omega}{\omega_\mathrm{D}})(\hat{d}_+^\dagger\hat{d}_-+\hat{d}_+\hat{d}_-^\dagger )+......,
    \end{split}
\end{equation}
where the left out term is non-resonant. Obviously, we see the $\hat{d}_-$ mode is coupled to $\hat{d}_+$ negative one side-band with coupling strength $g_{cm}J_1(\Omega/\omega_\mathrm{D})$, which leads to the anti-crossing in the reflection spectrum. In the experiment, we didn't see any anti-crossing between the $\pm 1$ side-band of each hybridized modes. The reason is obvious if we notice that there is only odd terms in the summation. Specifically, for example, there is no $n=2$ term, which means in whatever rotating frame we can't get a Hamiltonian in the form
\begin{equation}
\begin{split}
    \hat{H}&=(\omega_+-\omega_\mathrm{D})\hat{d}_+^\dagger\hat{d}_++(\omega_-+\omega_\mathrm{D})\hat{d}_-^\dagger\hat{d}_-\\&+{g_{cm}J_2(\frac{\Omega}{\omega_\mathrm{D}})(\hat{d}_+^\dagger\hat{d}_-+\hat{d}_+\hat{d}_-^\dagger )}+......
    \end{split}
\end{equation}
Thus there will be {no anti-crossing} between the two $\pm 1$ side bands. In fact, we can generalize the result: denote $k$ ($l$) as the $k_{th}$ ($l_{th}$) side-band of $\hat{d}_+$ ($\hat{d}_-$) mode, then there will be no anti-crossing between the side-bands if $k+l=even$.

\section*{Appendix IV: Autler Townes splitting}

Similar to a driven two level system, there is mode splitting when we apply a weak drive. This can be easily seen by rewriting Eq. \ref{eq1} in terms of the hybridized mode and express it in the rotating frame $U=\exp(-i\hat{d}_+^\dagger\hat{d}_+\varepsilon(t)/2-i\hat{d}^\dagger_-\hat{d}_-\varepsilon(t)/2)$, thus we have
\begin{equation}
    \hat{H}=\omega_+\hat{d}_+^\dagger\hat{d}_++\omega_-\hat{d}_-^\dagger\hat{d}_-+\frac{\Omega}{2}\cos(\omega_\mathrm{D}t)(\hat{d}_+^\dagger\hat{d}_-+\hat{d}_+\hat{d}_-^\dagger),
\end{equation}
where the hybridized mode frequency $\omega_+=\omega_0+g_{cm},\omega_-=\omega_0-g_{cm}$. We see that the problem becomes resembling a two level system driven by a laser field. The Autler Townes splitting just follows. A standard method of solving the Hamiltonian is to adopt the rotating wave approximation. First, take the Hamiltonian in the rotating frame, we have $U=\exp({i\omega_+\hat{d}^\dagger_+\hat{d}_+t+i\omega_-\hat{d}^\dagger_-\hat{d}_-t})$,
\begin{equation}
\begin{split}
    \hat{H}&=\frac{\Omega}{2} \cos(\omega_\mathrm{D}t)(\hat{d}_+^\dagger\hat{d}_-e^{i(\omega_+-\omega_-)t}+\hat{d}_+\hat{d}_-^\dagger e^{-i(\omega_+-\omega_-)t})\\
    &=\frac{\Omega}{2} (\frac{e^{i\omega_\mathrm{D}t}+e^{-i\omega_\mathrm{D}t}}{2})(\hat{d}_+^\dagger\hat{d}_-e^{i\Delta\omega t}+\hat{d}_+\hat{d}_-^\dagger e^{-i\Delta\omega t})\\
    &\simeq \frac{\Omega}{4} (\hat{d}_+^\dagger\hat{d}_-e^{i(\Delta\omega-\omega_\mathrm{D}) t}+\hat{d}_+\hat{d}_-^\dagger e^{-i(\Delta\omega-\omega_\mathrm{D}) t}),
\end{split}    
\end{equation}
where we define $\Delta\omega=\omega_+-\omega_-$, and the counter rotating term is ignored in the third line. We can further write down an equivalent form
\begin{equation}
    \hat{H}=\frac{\delta\omega_\mathrm{D}}{2}\hat{d}^\dagger_1\hat{d}_+-\frac{\delta\omega_\mathrm{D}}{2}\hat{d}_-^\dagger\hat{d}_-+\frac{\Omega}{4}(\hat{d}_+^\dagger\hat{d}_-+\hat{d}_+\hat{d}_-^\dagger),
\end{equation}
where $\delta\omega_\mathrm{D}=\Delta\omega-\omega_\mathrm{D}$. We thus obtain a time-independent Hamiltonian, and it can be easily diagonalized. Each mode furthre splits into two branches with the separation given by
\begin{equation}
    \Delta=\sqrt{\frac{\Omega^2}{4}+\delta_{\omega_\mathrm{D}}^2}.
\end{equation}

\section*{Appendix V: Model comparison}
\begin{figure}[tb]
	\centering
	\includegraphics[width=\linewidth]{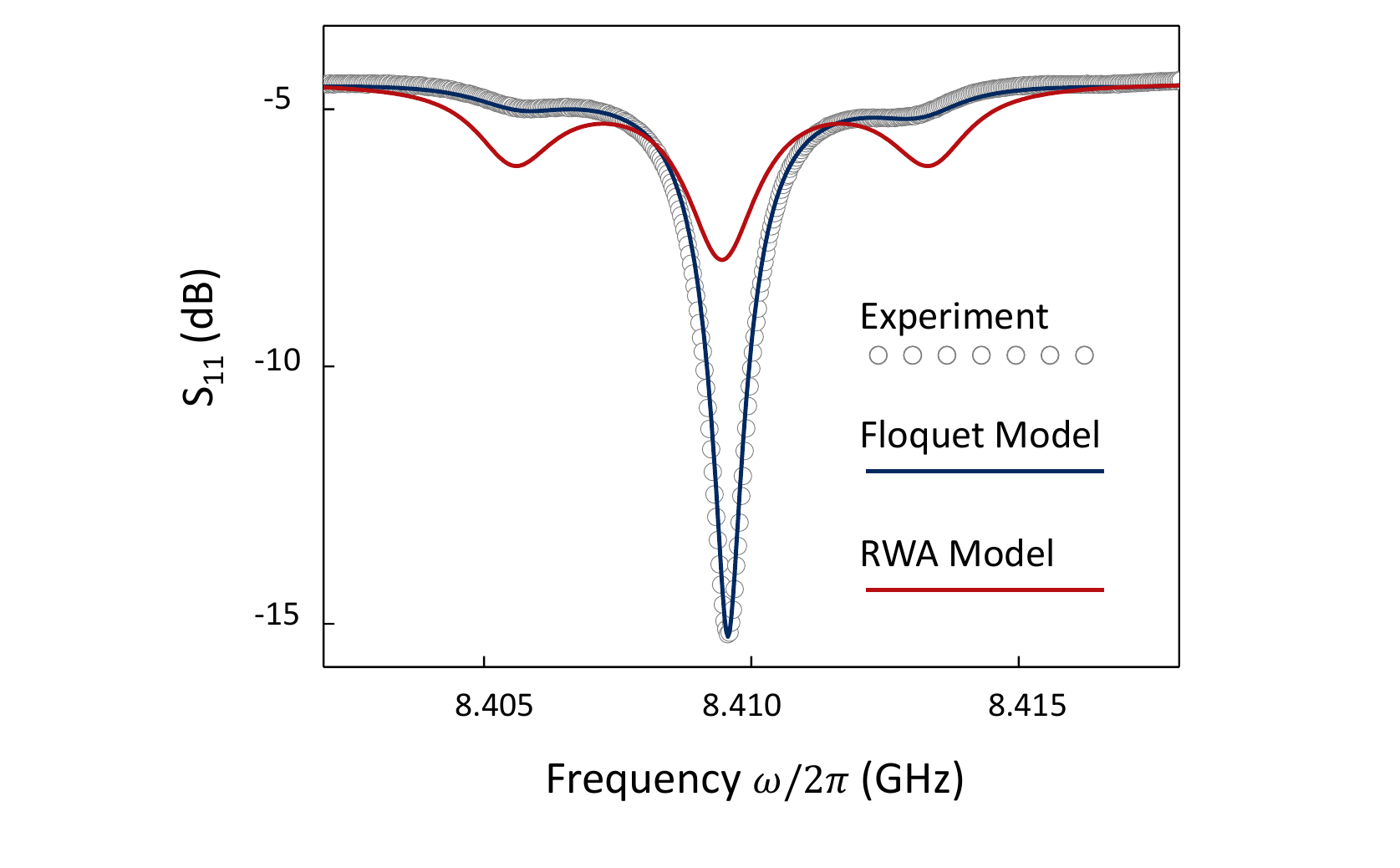}
	\caption{Line plots comparing the experiment and calculation results for $\omega_\mathrm{D}=2\pi\times 3.85$ MHz and a driving amplitude of 10 V peak-to-peak, which corresponds to $\Omega=2\pi\times8$ MHz.}
	\label{figS1}
\end{figure}

Figure\,\ref{figS1} provides a more detailed comparison of the calculation results using the Floquet model and the RWA model versus the measurement result. It is evident that the Floquet model shows excellent agreement with the experimental results while the RWA model shows significant deviation. These line plots correspond to the maximum driving amplitude condition in the intensity maps of Figs.4 (d)--(f) in the main text.

\section*{Appendix VI: System configuration and parameters}
The dielectric resonator we used has a diameter of 6 mm and height of 4 mm. It is hosted inside a copper housing to eliminate radiation losses. The housing is sufficiently large (diameter: 32 mm; height: 25 mm) to minimize the absorption of the metal wall. A Teflon spacer (not shown in Fig.\,\ref{figS2}a) is used to support the cylinder inside the housing. TE$_{01\delta}$ is the fundamental mode of this dielectric resonator. The loop antennas are made by terminating non-magnetic, semi-rigid coaxial cables with a loop. The innder diameter of the driving loop is around 900 $\mu$m, which is sufficiently small to enhance the driving efficiency but not too small to severely perturb the magnon resonance. The YIG sphere is glued inside the loop of the driving antenna using curable varnish. It has a diameter of 400 $\mu$m with a highly polished surface. The sphere is made of high-quality single-crystal YIG, with a saturation magnetization of 1780 Oe. 

System parameters extracted from experimental results in Figs.\,1 and 2 of the main text are: cavity resonance frequency $\omega_c=2\pi\times 8.505$ GHz; coupling strength between cavity resonance and fundamental magnon mode ($m_0$) $g_{cm0}=2\pi\times 14.0$ MHz; coupling strength between cavity resonance and the second magnon mode ($m_1$) $g_{cm1}=2\pi\times 6.3$ MHz; dissipation rate of fundamental magnon mode $\kappa_{m0}=2\pi\times4.4$ MHz; dissipation rate of second magnon mode $\kappa_{m1}=2\pi\times3.7$ MHz; dissipation rate of cavity photon mode $\kappa_c=2\pi\times2.0$ MHz. In Fig.\,\ref{figS2}, the cavity resonance is shifted to $\omega_c=2\pi\times 8.41$ GHz with a dissipation rate: $\kappa_c=2\pi\times1.5$ MHz. With the maximum driving amplitude that is available in our measurements (10 V peak-to-peak), the corresponding driving strength is calculated as $\Omega=2\pi\times 8$ MHz. Fundamental magnon mode ($m_0$) is used in Fig.\,2 and Fig.\,3 in the main text. The second magnon mode ($m_1$) is used in Fig.\,\ref{figS2}.

\section*{Appendix VII: Magnetic Tunability}
\begin{figure}[tb]
	\centering
	\includegraphics[width=\linewidth]{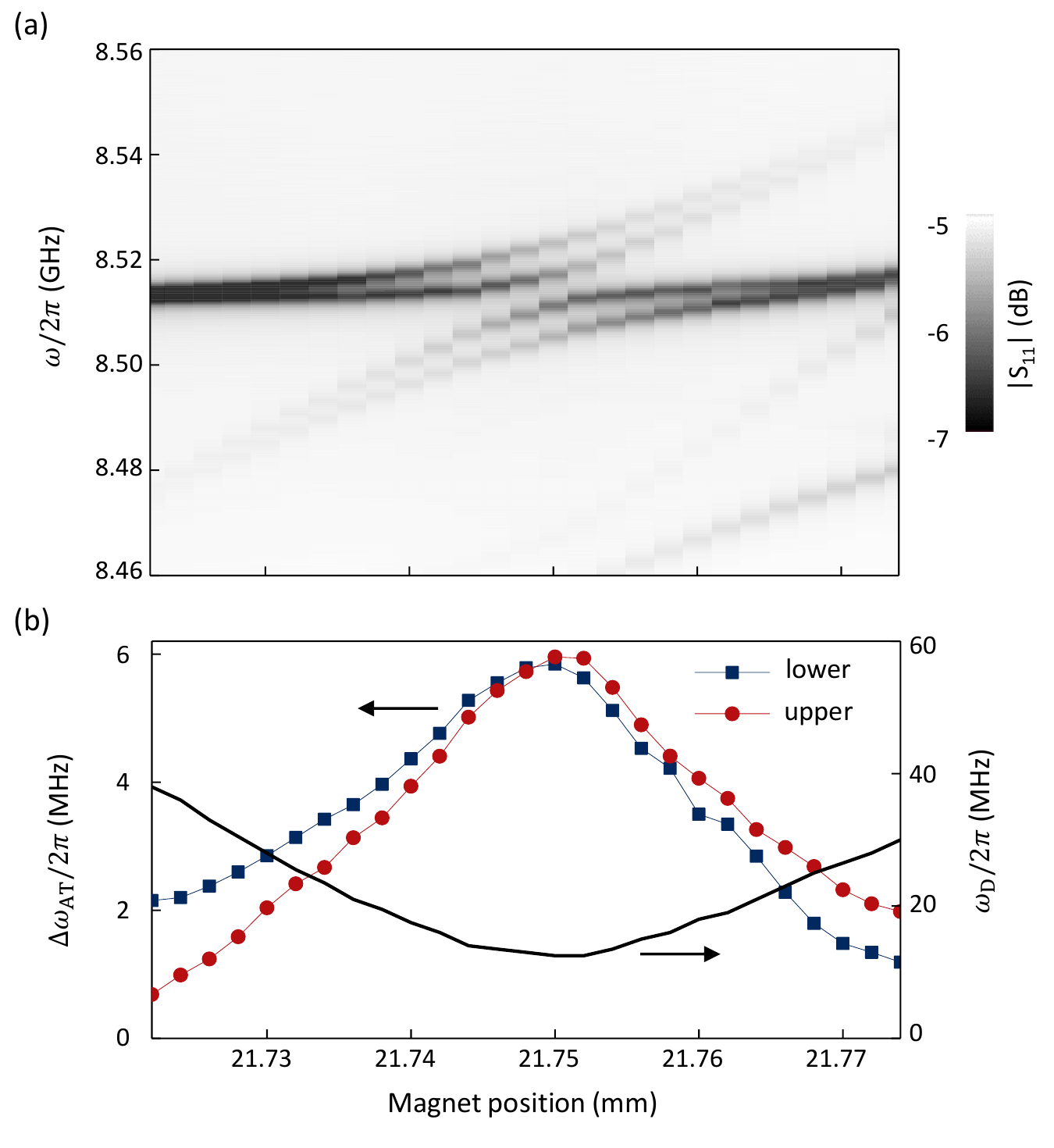}
	\caption{\textbf{a}, Summary of the cavity reflection spectra measured with $\omega_\mathrm{D}=\Delta\omega$ for different magnet positions. The driving amplitude is 10 V peak-to-peak. \textbf{b}, Extracted Autler-Townes splitting (ATS) $\Delta\omega_\mathrm{AT}$ for both the lower and upper branches (left y axis), and driving frequency $\omega_\mathrm{D}$ needed to observing ATS (right y axis) as a function of the magnet position.}
	\label{figS2}
\end{figure}

One distinct advantage of cavity electromagnonics is that the magnon frequency can be conveniently tuned by the external field. This allows us to study the Floquet driven magnon-photon coupling under various detuning conditions. The most prominent effect of the magnetic field is that it determines the driving frequency for ATS. From the above zero-detuning analysis, it is clear that ATS only occurs when the driving frequency matches the level separation: $\omega_\mathrm{D}=\Delta\omega$. However, the detuning can take non-zero values in our system and it is therefore magnetically tunable, which accordingly affects the level separation $\Delta\omega=\sqrt{4g_{cm}^2+\delta\omega^2}$ and consequently the required driving frequency $\omega_\mathrm{D}$. Figure\,\ref{figS2}a plots the measured cavity reflection spectra with ATS at various bias magnetic fields. The driving frequency needed for observing ATS is summarized in Fig.\,\ref{figS2}b as a function of the bias field.

In addition, the splitting ($\Delta\omega_\mathrm{AT}$) of each hybrid mode $d_\pm$  is also controlled by the bias magnetic field, as shown in Fig.\,\ref{figS2}a. This is more clearly shown in Fig.\,\ref{figS2}b, where the splitting varies with the bias field for both hybrid modes and reaches the maximum when magnon and cavity photon modes are on resonance (at $x=21.75$ mm). These capability of off-resonance operations demonstrated here enables new approaches for hybridizing magnons and microwave photons. For instance, magnons can be applied off-resonance to have minimal intrinsic coupling with cavity photons, and then the magnon-photon coupling can be achieved by the Floquet drive in a controllable way which allows the coupling to be completely turned off and on. This is different from the on-resonance operations where even if the Floquet drive is turned off, magnons and microwave photons are still hybridized because of the intrinsic coupling.

	
%
	
\end{document}